# Direct measurement of the electron density of extended femtosecond laser pulse-induced filaments

Y.-H. Chen, S. Varma, T. M. Antonsen, and H.M. Milchberg

*Institute for Research in Electronics and Applied Physics*

*University of Maryland, College Park, MD 20742*

Abstract

We present direct time- and space- resolved measurements of the electron density of femtosecond laser pulse-induced plasma filaments. The dominant nonlinearity responsible for extended atmospheric filaments is shown to be field-induced rotation of air molecules.

Long range filamentary propagation of intense femtosecond laser pulses in gases occurs due to the dynamic balance between nonlinear self-focusing of an intense optical pulse and laser plasma-induced defocusing. The filament formation process is initiated when the electric field strength of the propagating laser pulse is sufficient to nonlinearly distort the electron clouds of the gas atoms and molecules as well as rotate the molecules, leading to an ensemble averaged dipole moment nonlinearly increasing with field strength. This is manifested as an effective refractive index which acts to focus the beam, leading to plasma generation when the gas ionization intensity threshold is exceeded. The on-axis concentration of free electrons then defocuses the beam. This dynamic balancing between self-focusing and ionization-defocusing leads to self-sustained propagation accompanied by electron density tracks over distances from centimeters to hundreds of meters in atmosphere. Long-range filamentation by femtosecond pulses in gases was first observed in 1995 [1] and has been subsequently studied by many groups pursuing both basic understanding and applications [2].

Plasma generation is the key to extended filamentary propagation. In this Letter we present for the first time direct time-resolved measurements of electron density profiles along and across the path of the filamenting pulse. These measurements allow us to clearly identify the molecular rotational response of nitrogen and oxygen as the dominant contribution to extended atmospheric filament dynamics.

Given the centrality of the electron density profile to the physics and applications of femtosecond filaments, the lack to date of its direct space and time resolved measurement is striking. Prior estimates of electron density and filament width have been obtained through simulations [3-5] and through several indirect experimental methods including optical probing [6,7], plasma fluorescence [7], and secondary electrical discharges [8]. For filamentation formation in atmosphere, theory and simulation have predicted typical peak electron densities anywhere from $\sim 10^{12}$ cm$^{-3}$ through $10^{17}$ cm$^{-3}$ depending on laser and focusing conditions. Experiments have also produced estimates in this range, although lack of spatial resolution has made it unclear whether single or multiple filaments were being measured.

Figure 1(a) shows our experimental geometry. A rail-mounted folded wavefront optical interferometer can travel ~2 meters parallel to the filament axis. A probe beam is split from the main 800 nm filamentation beam, counter-propagated at an angle $\theta=0.75°$ across the filament, and sent through the interferometer. A plane central to the crossing region is imaged with a microscope objective. The variable probe delay allows variable lengths of filament to be imaged. For our typical measured filament diameters here of $d_{fil}$ ~70 µm, the probe path through the plasma is $d_{probe} \sim d_{fil}/\sin(\theta)$~5mm, the axial spatial resolution along the filament is $\Delta z_{res} \sim d_{fil}/\tan(\theta)$ ~5 mm, the temporal resolution is $\Delta t_{res} = \Delta z_{res}/c$ ~10-15 ps, and the radial resolution is 5 µm. Crucial to our ability to extract the very small optical phase shifts imposed by the low filament electron densities is (*i*) the extended probe interaction length $\Delta z_{res}$ and (*ii*) a very high quality probe phase front imposed by a spatial filter. We have verified that $d_{probe}$ is sufficiently short for our filament densities that negligible refractive distortion of the probe phase front occurs. Our resolution limit is set by residual phase front noise in the probe which sets our measurable lower bound density to ~ $5 \times 10^{14}$ cm$^{-3}$. Both phase and electron density are extracted from the raw interferograms using standard techniques [9].

As has been well-visualized by simulations [2, 3-5], a femtosecond filament in gas starts as a result of rapid self-focusing instability arrested by plasma formation and defocusing. We have observed this abrupt beam collapse and electron density onset as shown in Fig. 1(b), for the case of a 2.85 mJ, 72 fs pulse (40 GW) focused at f-number =$f_\#$=345 (f=224cm, beam diameter=6.5mm) using a probe delay $\Delta t_{probe}$= 50 ps. The frames show a sequence of 1 cm spaced axial positions of the interferometer object plane, (i) z= −39cm through (vi) −34cm, where z=0 is the lens focal plane. It is first worth explaining the characteristic 'bowtie' shape of the phase images. These reflect the short (~1 mm) depth of field of the imaging system: the wide sections are out-of-focus regions on either side of the object plane, which is tightly imaged at the bowtie centre. The local filament axial and radial density profile (see Fig. 2) is obtained from phase extraction only at the centre of the bowtie. In frame (i) of Fig. 1(b), at z=-39cm, only the right side of the bowtie is visible; there is no filament electron density at the object plane and upstream of it. Inspection of the object plane and upstream region in frames (ii) nd (iii) shows the abrupt onset of ionization at z>-38cm, while frames (iv)-(vi), looking back along the filament at increasing distances, show the filament's continued upstream development. The terminated right side of the bowtie is the leading temporal edge of the filament; the interferometer probe temporal delay was here adjusted to catch the filament mid-flight. Both far downstream from the collapse point and at longer probe delays, the bowtie extends and widens to both the left and right edges of the frame.

The ability to directly measure the electron density with good axial resolution allows a sensitive test of filament propagation physics. One of the most discussed aspects of filamentation has been the nature of the neutral gas nonlinearity first leading to beam collapse and later contributing to the dynamic stabilization. Part of the air nonlinearity is an instantaneous response owing to electron cloud nonlinear distortion in argon atoms and within $N_2$ and $O_2$ air molecules. It has also long been recognized [10] that molecular rotation in the laser field contributes a delayed nonlinearity resulting from the increased linear dipole moment as the molecular axis is torqued toward the laser polarization. Recently, it has been suggested that the orientational effect is in fact dominant at the typical ~100 fs pulse lengths and ~mJ energy levels used for a majority of air filamentation experiments [9, 11]. This is because the filamenting pulse, in the intensity range $10^{13}$ -$10^{14}$ W/cm$^2$, typically excites ~20-30 rotational quantum states in $N_2$ and $O_2$, resulting in a refractive index response time due to rotation alone of $\delta t_{rot}$ ~ $2T/j_{max}(j_{max}+1)$~ 20-40 fs [9, 11], where $T$ is the fundamental molecular rotational period ($T_{N2}$=8.3ps for nitrogen and $T_{O2}$ =11.6 ps for oxygen [12]). On the other hand, the instantaneous nonlinear response dominates the rotational response in the small molecules $H_2$ and $D_2$, as shown in [9].

To illustrate the profound effect of changes in the laser pulsewidth on air filamentation, we performed experiments with two different focusing geometries, keeping the peak laser power fixed for each: (a) lens focal length f= 95 cm, f-number= $f_\#$ =240, laser peak power $P$=17 GW, $\tau_{short}$=40 fs, $\tau_{long}$ =120 fs, and (b) f=306 cm, $f_\#$=505, P=19 GW, $\tau_{short}$ =44 fs, $\tau_{long}$=132 fs. Here, $\tau$ refers to the FWHM pulsewidths measured immediately after the lens by a field envelope/phase diagnostic [13] and a single shot autocorrelator. The pulsewidth was adjusted by translating the laser's compressor grating, and we verified that this amount of chirp had no effect on the results. In all experiments, the power was carefully adjusted to avoid development of hotspots in the beam prior to the lens, or multiple filamentation as seen by multiple white light spots on a screen in the far field. The interferometer probe delay was set to $\Delta t_{probe}$= 1 ps (we measured the density to rapidly drop in the first 100ps). We observed that the length, structure, and density of measured filaments is quite sensitive to minor distortions in the beam, either from optics or from hotspots due to accumulation of nonlinear phase in air in advance of the lens. A clean beam is essential for comparing experiments to propagation simulations. Interferometric phase images, from which the electron density profile was extracted, were averaged over 200 shots at each axial position. To compensate for slight shot-to-shot transverse movement of the filament, the phase images for each shot were spatially aligned by an image autocorrelation technique before averaging.

Figure 2(a) shows the on-axis filament electron density (left panel) $n_e$ and FWHM diameter $d_{fil}$ (right panel) as a function of axial position, corresponding to $f_\#$~ 240 and conditions (a) above. Peak $n_e$ ~6.5x10$^{16}$ cm$^{-3}$ (corresponding to fractional ionization $\eta$ ~2.4x10$^{-3}$) occurs at z ~ −9 cm for $\tau_{long}$, while peak $n_e$ ~2.4x10$^{16}$ cm$^{-3}$ ($\eta$ ~10$^{-3}$) occurs at z~ −3 cm for $\tau_{short}$, with $\tau_{long}$ resulting in stronger self-focusing and more rapid collapse, higher peak density, and longer overall filament length. For both pulses, $d_{fil}$ is quite stable over the filament length, except near the end. A notable difference is a secondary electron density peak at z~5cm for $\tau_{long}$. Both filaments are quite different from the case of propagation in very low air density, where a simulation (described below) for $\tau_{short}$ shows an axial density distribution peaking at $\eta = 10^{-5}$, with an axial FWHM of ~10 cm centered at z=0. For $\tau_{long}$, the low density simulation shows an even smaller $\eta$ and axial extent.

For $f_\#\sim505$ and conditions (b) above, Fig. 2(b) shows an earlier axial beam collapse and filament onset for $\tau_{long}$ than for $\tau_{short}$, with the entire measurable filament located in advance of z=0 in both cases. The peak density for $\tau_{long}$ is $3.6\times10^{16}$ cm$^{-3}$ ($\eta\sim10^{-3}$), while that for $\tau_{short}$ is $1.1\times10^{16}$ cm$^{-3}$. $d_{fil}$ for both pulses is in the range ~65–80 μm, with widening near the end of the measurable filament. Here, as in the small $f_\#$ case, $\tau_{long}$ results in stronger self-focusing, higher peak density, longer overall filament length, and the appearance of a prominent secondary density peak. By contrast, low density simulations for these conditions show $\eta < 10^{-7}$ centered at z=0.

Figure 3 compares filament output on-axis spectra for $\tau_{long}$ and $\tau_{short}$ for the two focusing cases. The input spectra for the long and short pulses are shown as thinner curves. In all cases the output spectrum for $\tau_{long}$ is wider and more intense, owing to the greater effective nonlinearity and extended interaction length experienced by that pulse. For the $f_\#\sim505$ filament of panel (ii), the enhanced spectral broadening for $\tau_{long}$ is particularly strong, and images of the filament white light spots are shown as insets.

We have simulated intense femtosecond pulse propagation in air using our code WAKE [5, 14], assuming a cylindrically symmetric, extended paraxial wave equation for the laser pulse, $\frac{2}{c}\frac{\partial}{\partial s}\left(i\omega_0 + \frac{\partial}{\partial \tau}\right)a - \beta_2\frac{\partial^2 a}{\partial \tau^2} + \nabla_\perp^2 a = \left(\frac{4\pi q^2 n_e}{mc^2} - k_0^2 \delta\varepsilon\right)a$. Here, $a(\mathbf{x}_\perp,\tau,s)$ is the complex envelope of the laser electric field, $k_0$ and $\omega_0$ are the central wavenumber and frequency of the initial pulse, $\mathbf{x}_\perp$ is the transverse coordinate, $s=z$ is propagation distance, $\tau = t - z/v_g$ is time local to the pulse frame moving at group velocity $v_g$, $\beta_2 = k_0(d^2k/d\omega^2)_{\omega=\omega_0}$ is the group velocity dispersion of air, $q=-e$ is the electron charge, $n_e$ is electron density generated by laser ionization, and $\delta\varepsilon(\mathbf{x}_\perp,\tau) = \delta\varepsilon_{rot} + \delta\varepsilon_{inst}$ is the nonlinear response. $\delta\varepsilon_{rot}$ is modeled as a damped oscillator

$$[\frac{d^2}{d\tau^2} + 2\gamma_D \frac{d}{d\tau} + \omega_m^2]\delta\varepsilon_{rot}(\mathbf{x}_\perp,\tau) = 2\omega_m^2 n_{20} I(\mathbf{x}_\perp,\tau)$$ with $I(\mathbf{x}_\perp,\tau) \propto |a|^2$, and where $2\pi/\omega_m =$ 400fs, $\gamma_D = 120$ fs, and $n_{20} = 2.6 \times 10^{-19}$ cm$^2$/W are obtained from fits to our prior ultrafast measurements of weighted $N_2$ and $O_2$ rotational response [9]. Ignored are later-time rotational quantum revivals [11]. In our prior measurements [9], comparatively negligible instantaneous response in $N_2$ and $O_2$ was observed for $\tau < \delta t_{rot}$, and here we use $\delta\varepsilon_{inst}^{max} / \delta\varepsilon_{rot}^{max} < 0.15$ as an upper bound setting the value of $n_2$ in $\delta\varepsilon_{inst} = 2n_2 I(\mathbf{x}_\perp,\tau)$, Therefore, at long times $\tau >> 2\pi/\omega_m$, $\delta\varepsilon(\mathbf{x}_\perp,\tau) \sim 2(n_{20}+n_2)I(\mathbf{x}_\perp,\tau)$, where our value for $n_{20}+n_2 \sim 3.0 \times 10^{-19}$ cm$^2$/W corresponds closely with measured "$n_2$" values for long pulses [2]. Our apertured beam is modeled as a flat top with a smooth intensity transition to zero at 90% radius, while the phase was taken as a quadratic function of radius determined by the lens focal length. A fit to tunneling/multiphoton rates [15] was used to simulate air ionization, as this rate is higher than tunneling alone in the expected intensity range of mid-$10^{13}$ W/cm$^2$. For our parameters, the paraxial approximation is sufficient.

Simulation results for on-axis filament density and filament diameter vs. axial position are shown in Fig. 4 for the focusing geometries, peak powers and pulsewidths of Fig. 2, plus additional runs at 20% higher power. The additional runs were performed because for $f_\# \sim 505$ and $\tau_{short}$, experimental filament onset was quite sensitive to energy: for the lower power case for $\tau_{short}$, it is seen that the peak filament density drops to almost half.

The main qualitative features of the simulations are in agreement with the measurements. *For the same peak power*, the $\tau_{long}$ filaments start earlier and more abruptly, have significantly higher density, and are longer than those for $\tau_{short}$. As well, the $\tau_{long}$ filaments have multiple peaks as in the experiments. The absolute density values and filament extents are in good agreement for the $f_\#=240$ case. For $f_\#=505$, while the peak densities differ by ~50%, the qualitative behaviour is well-followed. We note that varying the simulation's ionization model varies the detailed shape of the radial electron density profile (but not the peak density; see further discussion below). A specific measure of width (such as FWHM) can then show large variation for different ionization models used. The need to refine the ionization model may explain the difference between experimental and simulated filament FWHM.

The difference in long and short pulse results is a direct consequence of the dominance of the delayed nonlinear response of the air molecules, which leads to stronger nonlinear focusing and more extended propagation. The double electron density peaks originate from the variation of nonlinear focusing and refraction through the temporal envelope of the pulse: the leading portion loses energy and refracts away from its self-generated plasma, causing a dip in the electron density while the trailing part accumulates sufficient nonlinear phase in the filament periphery to self-focus and cause a plasma resurgence [3-5]. In air, this effect is enhanced for longer pulses since the later slices of the pulse experience increasing molecular rotational nonlinearity. The multiple density peaks are associated with pulse temporal splitting, which we have measured on-axis and will present in a future publication. The higher electron density for $\tau_{long}$ is simply the filament's dynamical offset of the stronger nonlinear focus of the excited molecular lens. In effect, one can control density/intensity clamping in atmosphere by controlling the molecular excitation. The simulation's sensitivity to ionization model was tested by multiplying the rate by a factor of 20. This resulted in negligibly increased peak densities, reinforcing the idea that the density and intensity are clamped by a dynamic balance of plasma-induced refraction and self focusing. Finally, we note that our laser and focusing parameters for $\tau_{short}$ (red circles) in Fig. 2(a) correspond closely to parameters simulated in ref. [16]. Using a speculative model for the air nonlinearity, ref. [16] claims filamentation takes place with negligible ionization. Our experimental and simulation results in Fig. 2(a) and Fig. 4(a) say otherwise.

In conclusion, we have presented the first direct space and time resolved electron density measurements for a nonrelativistic femtosecond laser pulse nonlinearly filamenting in gas, here the atmosphere. This has allowed a detailed elucidation of the nonlinear physics leading to atmospheric filament formation and a route to its enhancement.

The authors thank A.B. Fallah and Yu-Hsiang Cheng for technical assistance, and acknowledge the support of the Office of Naval Research, the National Science Foundation, the US Dept. of Energy, and the Lockheed-Martin Corporation.

**References**

1. A. Braun *et al.*, Opt. Lett. **20**, 73 (1995).

2. For example, A. Couairon and A. Mysyrowicz. Physics Reports **441**, 47 (2007) and references therein.

3. M. Mlejnek, E. M. Wright, and J. V. Moloney, Opt. Lett. **23**, 382 (1998).

4. A. Couairon and L. Bergé, Phys. Plasmas 7, 193 (2000).

5. J. Wu and T.M. Antonsen, Phys. Plasmas **10**, 2254 (2003).

6. S. Tzortzakis *et al.*, Opt. Comm. **181**, 123 (2000), G. Rodriguez *et al.*, J. Opt. Soc. Am. B **25**, 1988 (2008).

7. F. Théberge *et al.*, Phys. Rev. E **74**, 036406 (2006).

8. S. Eisenmann, A. Pukhov, and A. Zigler, Phys. Rev. Lett. **98**, 155002 (2007).

9. Y.-H. Chen *et al.*, Opt. Express **15**, 11341 (2007); Y.-H. Chen *et al.*, Opt. Express **15**, 7458 (2007).

10. E.T.J. Nibbering *et al.*, J. Opt. Soc. Am. B **14**, 650-660 (1997).

11. S. Varma, Y.-H. Chen, and H. M. Milchberg, Phys. Rev. Lett. **101**, 205001 (2008).

12. C.H. Lin *et al.*, Phys. Rev. A **13**, 813 (1976).

13. P. O'Shea *et al.*, Opt. Lett. **26**, 932 (2001).

14. A. P. Mora and T. M. Antonsen, Jr., Phys. Plasmas 4, 217 (1997).

15. A. Talebpour, J. Yang, S.L. Chin, Opt. Comm. **163**, 29 (1999).

16. P. Béjot et al, Phys. Rev. Lett. **104**, 103903 (2010).

**Figure captions**

**Figure 1(a):** Experimental interferometry setup, showing images lenses and microscope objective. The inset shows the filament width *d* and probe crossing angle *θ*.

**Figure 1(b):** Spatial sequence of bowtie phase images showing filament collapse, for a 2.85 mJ, 72 fs pulse focused at $f_\#$ =345. The frames show a sequence of 1 cm spaced axial positions of the interferometer object plane, (i) z=-39 cm through (vi) z= -34 cm.

**Figure 2:** On-axis electron density and filament FWHM vs. axial distance for (a) lens focal length f= 95 cm, $f_\#$ =240, laser peak power *P*=17 GW, $\tau_{short}$=40 fs (red circles), $\tau_{long}$=120 fs (black squares), and (b) f=306 cm, $f_\#$=505, *P*=19 GW, $\tau_{short}$ =44 fs (red circles), $\tau_{long}$=132 fs (black squares). Each point is a 200 shot average. The inset in (b) shows a sample electron density profile from z=−75cm. Peak values shown have ~20% uncertainty near the collapse point and ~10% over the rest of the filament.

**Figure 3:** On-axis filament spectra corresponding to the conditions of Fig. 2, with (i) $f_\#$=240 and (ii) $f_\#$=505. Also shown are the input spectra. The insets in (ii) show the filament spots on a far field screen for $\tau_{short}$ and $\tau_{long}$.

**Figure 4:** Simulation results for the conditions of Fig. 2 (solid lines). Additional runs have been performed at 20% higher power (dashed lines). Red- short pulse; Black-long pulse.

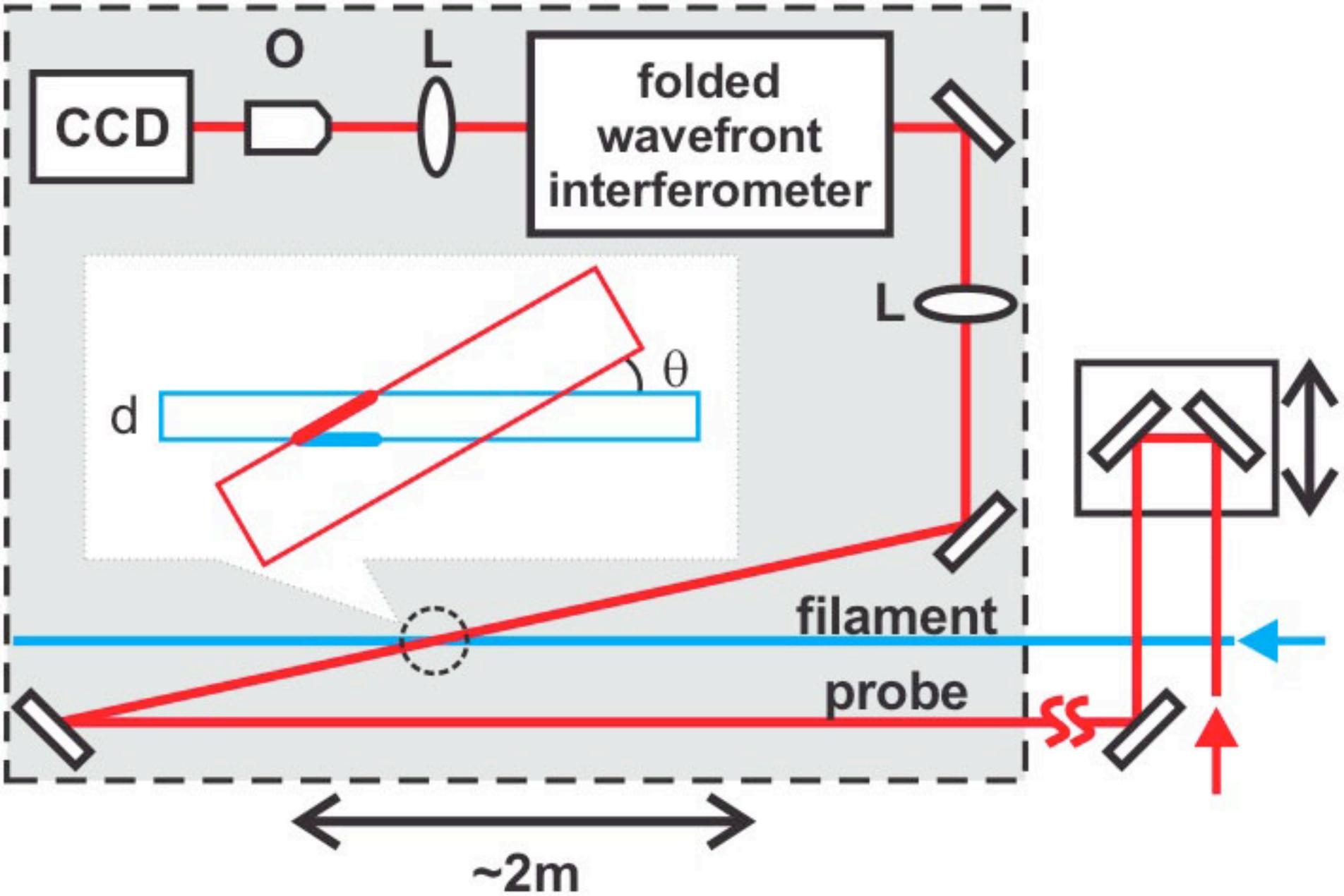

Fig. 1a

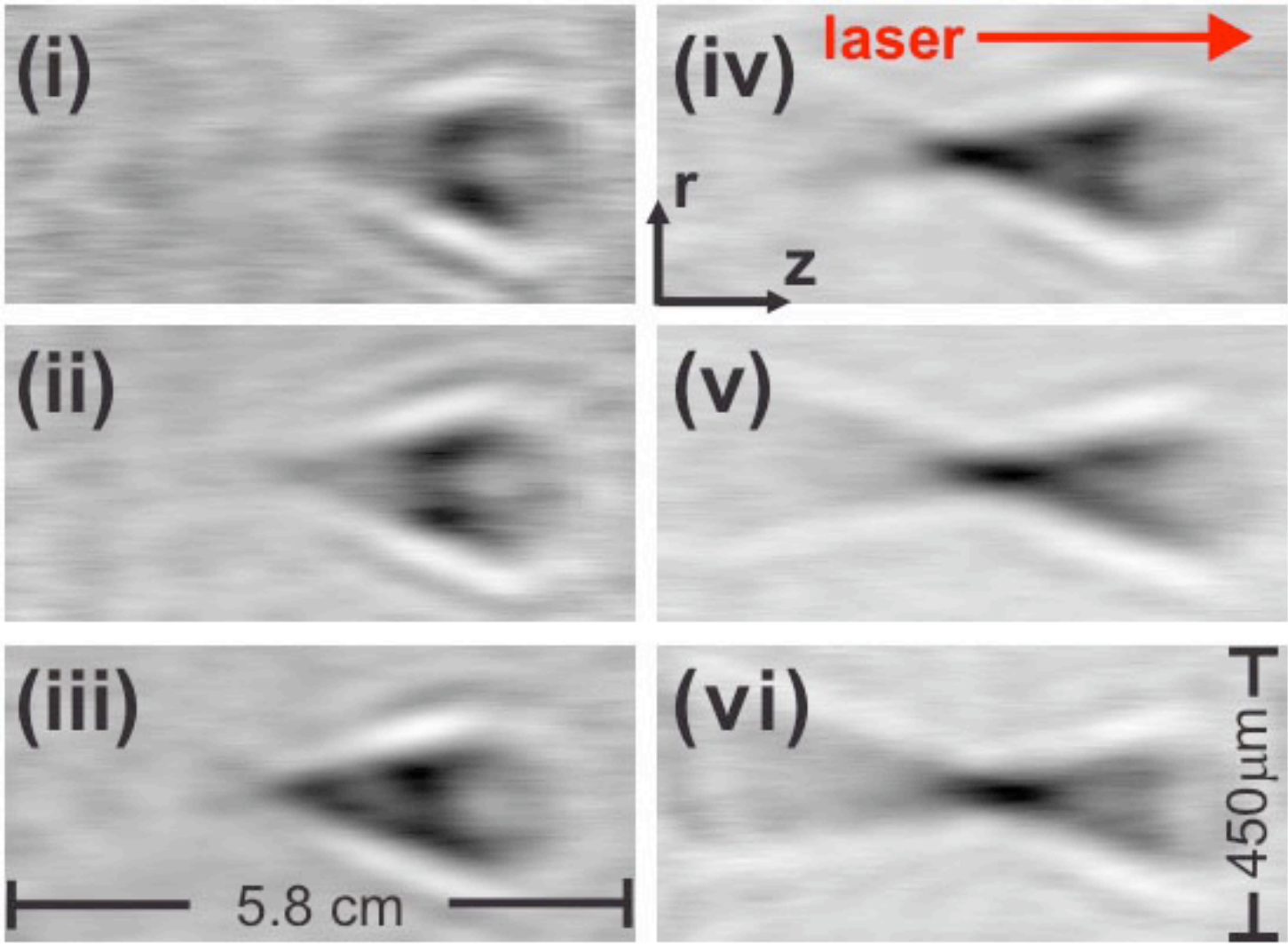

Fig. 1b

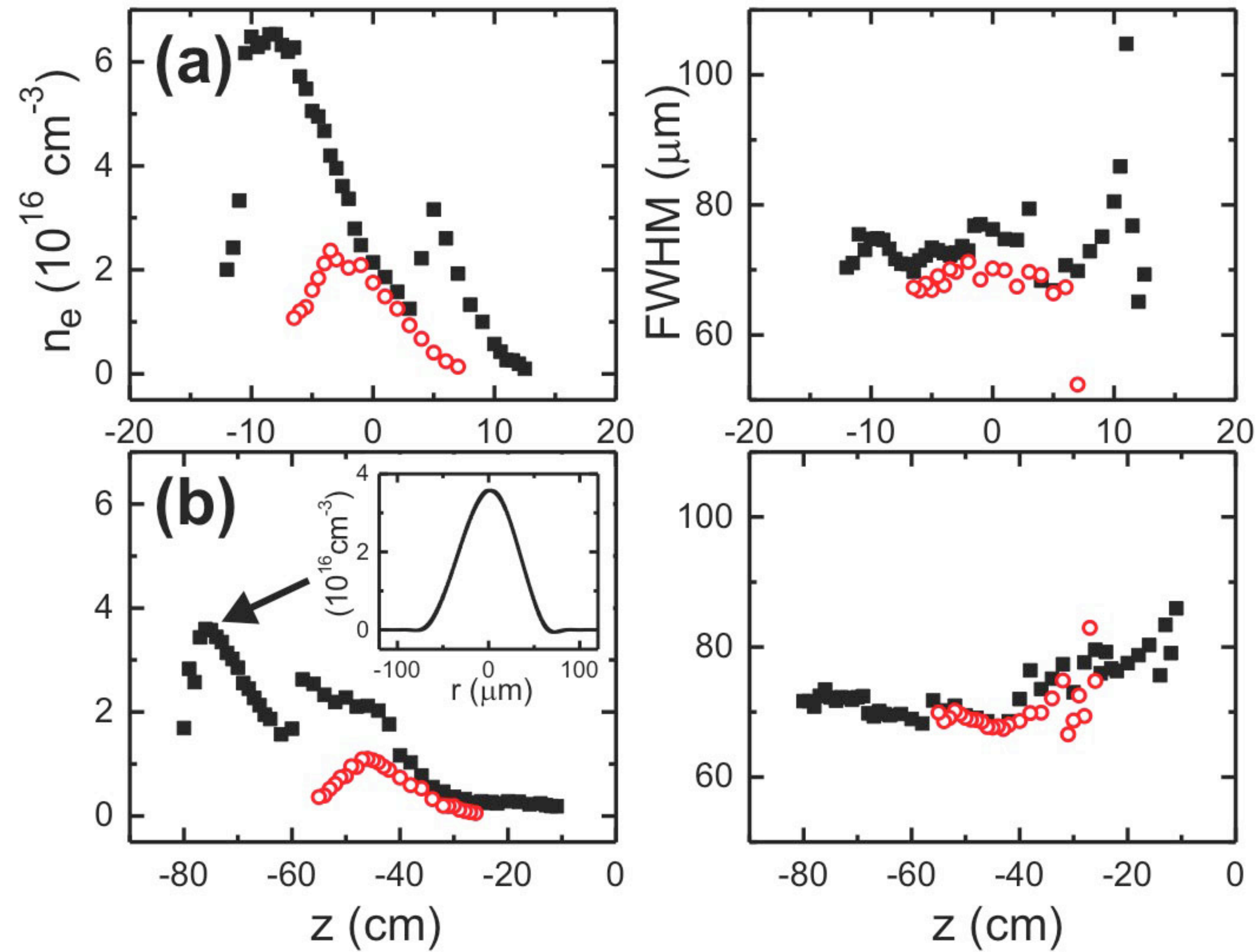

Fig. 2

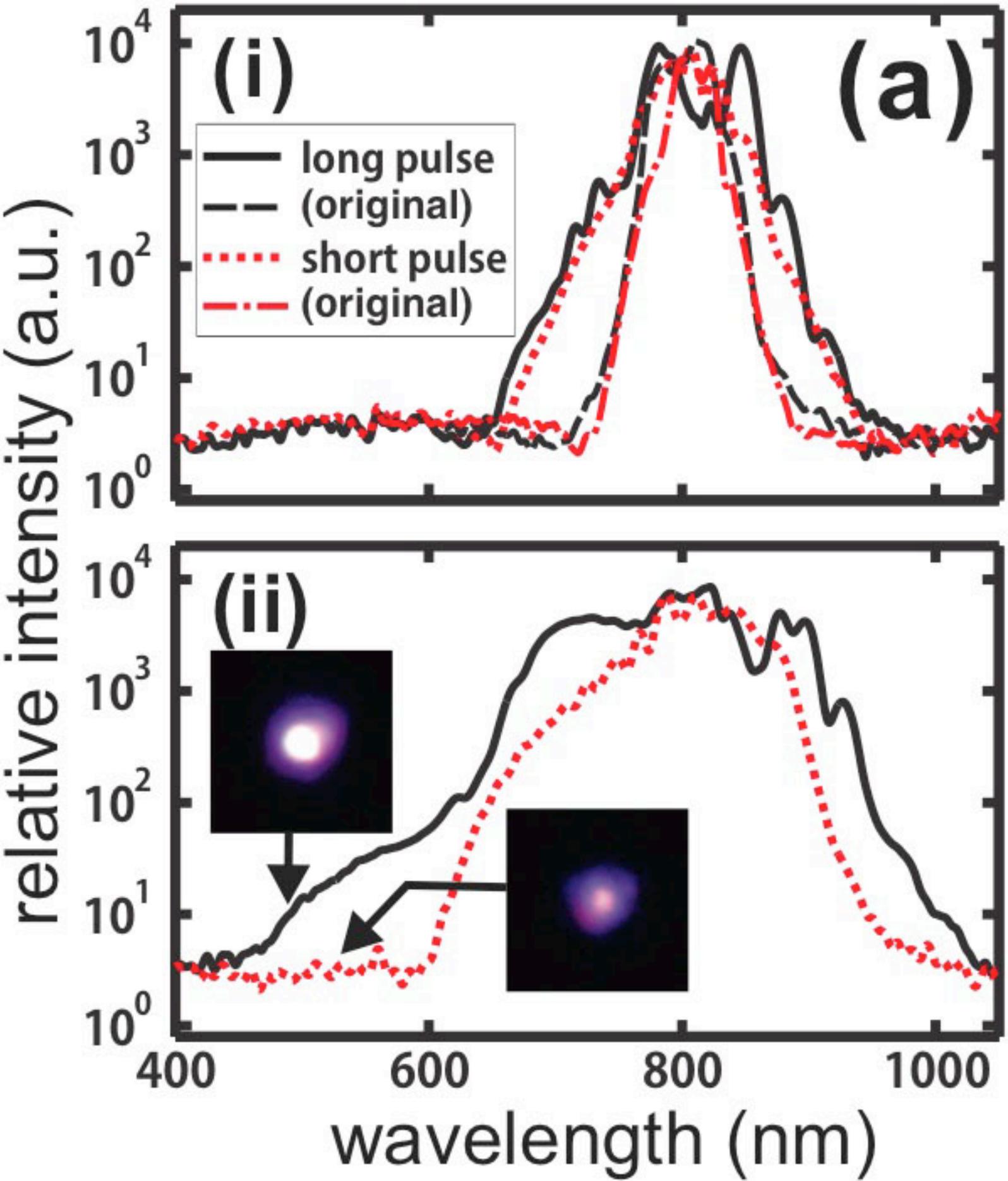

Fig. 3

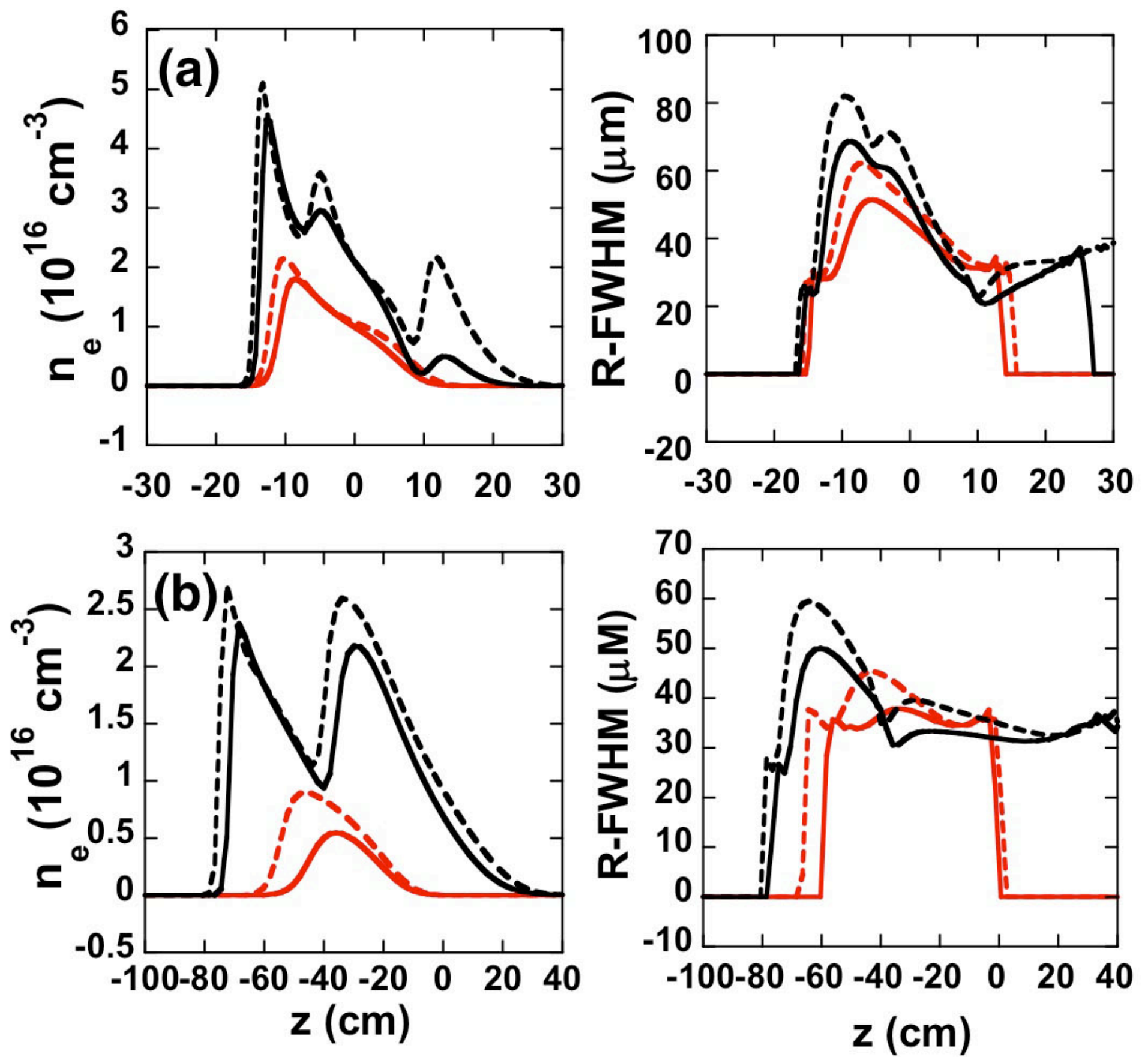

Fig. 4